\documentclass[journal,twoside,web]{ieeecolor}
\usepackage{generic, cite, amsmath, amssymb, amsfonts, algorithmic, graphicx, textcomp, threeparttable, xcolor, float, multirow, tikz, pifont, makecell, booktabs}
\usepackage[hidelinks]{hyperref}
\usepackage[outdir=./]{epstopdf}

\hypersetup{
    colorlinks=true,
    linkcolor=black,
    urlcolor=blue,
    }
\newcommand{\tabincell}[2]{\begin{tabular}{@{}#1@{}}#2\end{tabular}}

\def\BibTeX{{\rm B\kern-.05em{\sc i\kern-.025em b}\kern-.08em
    T\kern-.1667em\lower.7ex\hbox{E}\kern-.125emX}}
\DeclareRobustCommand{\orcidicon}{
\begin{tikzpicture}
\draw[lime, fill=lime] (0,0)
circle[radius=0.16]
node[white]{{\fontfamily{qag}\selectfont \tiny \.{I}D}};
\end{tikzpicture}
\hspace{-2mm}
}
\foreach \x in {A, ..., Z}{
\expandafter\xdef\csname orcid\x\endcsname{\noexpand\href{https://orcid.org/\csname orcidauthor\x\endcsname}{\noexpand\orcidicon}}
}

\begin{document}
%\title{A Sub-pixel Accurate Partial Image Phase-Only Correlation Algorithm for Rheumatoid Arthritis Imaging}
\title{A Sub-pixel Accurate Quantification of Joint Space Narrowing Progression in Rheumatoid Arthritis}
\author{Yafei Ou\hspace{-1.5mm}\orcidO{}, Prasoon Ambalathankandy\hspace{-1.5mm}\orcidP{}, Ryunosuke Furuya, Seiya Kawada, Tianyu Zeng, Yujie An, Tamotsu Kamishima\hspace{-1.5mm}\orcidK{}, Kenichi Tamura, and Masayuki Ikebe\hspace{-1.5mm}\orcidI{}
%\thanks{This work was supported in part by the Japan Society for the Promotion of Science (JSPS) Grants-in-Aid for Scientific Research (KAKENHI) under Grant 18H05288 and Grant 21K07611 and in part by JST SPRING under Grant JPMJSP2119. \textit{(Corresponding author: Masayuki Ikebe.)}}
\thanks{Yafei Ou, Prasoon Ambalathankandy, Ryunosuke Furuya and Seiya Kawada are with the Research Center For Integrated Quantum Electronics, Hokkaido University, Sapporo 060-0813, Japan, and also with the Graduate School of Information Science and Technology, Hokkaido University, Sapporo 060-0814, Japan.}
\thanks{Tianyu Zeng and Yujie An are with the Graduate School of Health Sciences, Hokkaido University, Sapporo 060-0812, Japan.}
\thanks{Tamotsu Kamishima is with Faculty of Health Sciences, Hokkaido University, Sapporo 060-0812, Japan.}
\thanks{Kenichi Tamura is with Department of Mechanical Engineering, College of Engineering, Nihon University, Koriyama 963-8642, Japan.}
\thanks{Masayuki Ikebe is with the Research Center For Integrated Quantum Electronics, Hokkaido University, Sapporo 060-0813, Japan (e-mail: ikebe@ist.hokudai.ac.jp).}
\thanks{This work has been published in IEEE Journal of Biomedical and Health Informatics. DOI: \href{https://ieeexplore.ieee.org/document/9931435}{10.1109/JBHI.2022.3217685}.}
}

\maketitle
\begin{abstract}
Rheumatoid arthritis (RA) is a chronic autoimmune disease that primarily affects peripheral synovial joints, like fingers, wrist and feet. Radiology plays a critical role in the diagnosis and monitoring of RA. Limited by the current spatial resolution of radiographic imaging, joint space narrowing (JSN) progression of RA with the same reason above can be less than one pixel per year with universal spatial resolution. Insensitive monitoring of JSN can hinder the radiologist/rheumatologist from making a proper and timely clinical judgment. In this paper, we propose a novel and sensitive method that we call partial image phase-only correlation which aims to automatically quantify JSN progression in the early stages of RA. The majority of the current literature utilizes the mean error, root-mean-square deviation and standard deviation to report the accuracy at pixel level. Our work measures JSN progression between a baseline and its follow-up finger joint images by using the phase spectrum in the frequency domain. Using this study, the mean error can be reduced to 0.0130mm when applied to phantom radiographs with ground truth, and 0.0519mm standard deviation for clinical radiography. With its sub-pixel accuracy far beyond manual measurement, we are optimistic that our work is promising for automatically quantifying JSN progression.

\end{abstract}

\section{INTRODUCTION}

Rheumatoid arthritis (RA) is a progressive, chronic autoimmune disease characterized by synovitis that can ultimately cause deformities and ankylosis in peripheral synovial joints and impair the movement and flexibility of digits, and as well as the patient's whole hand. 
The major radiographic changes on hand, wrist and feet joints are cartilage damage and bone destruction (like bone erosion and joint space narrowing (JSN)).
Those damage and destruction typically lead to painful joints, progressive joint destruction, deformity, followed by functional limitation and severe disability \cite{cush2010rheumatoid, young2000does}.
There are substantial evidence that RA can be managed in a low level of disease activity and clinical remission with disease-modifying antirheumatic drugs\cite{saag2008american, majithia2007rheumatoid}.
Early diagnosis by precise quantification of subtle radiographic changes is essential for successful treatment, as it can improve outcomes and effectively manage the progression of RA \cite{saag2008american, majithia2007rheumatoid}.

\begin{figure}[!t]
\centering\includegraphics[width=\linewidth]{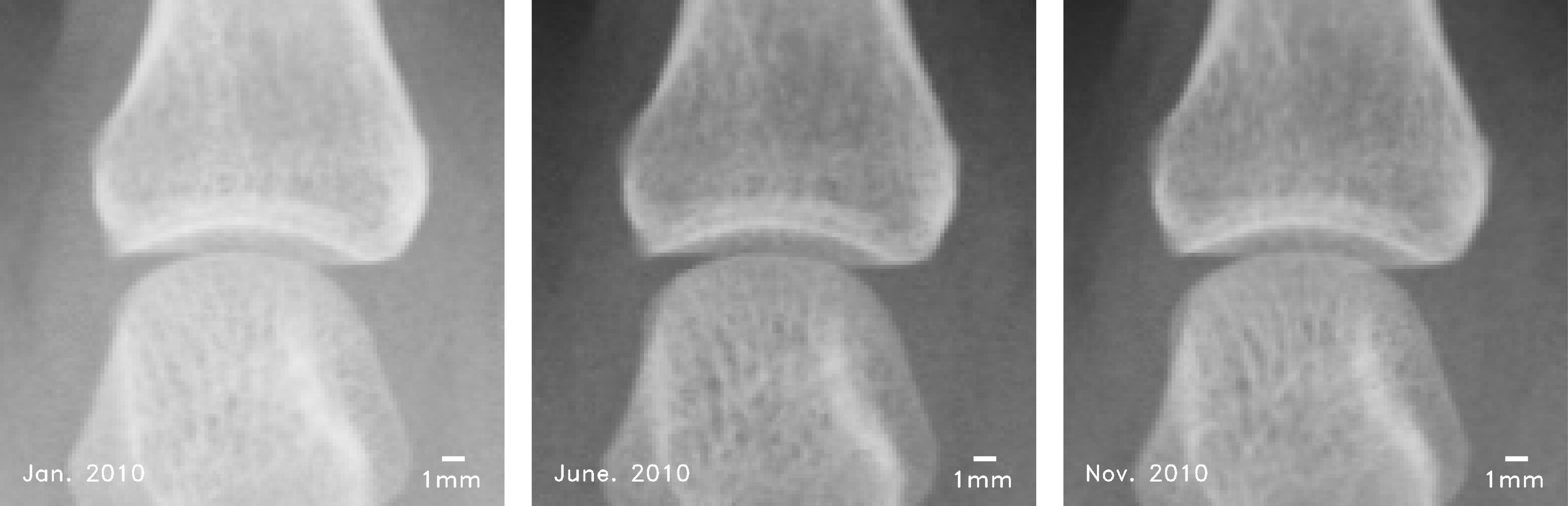}
\caption{JSN progression of a MCP joint for little finger over a period of 10 months. From left to right the images are: baseline, five-month, and ten-month images (spatial resolution: $0.175mm/pixel$). Usually, JSN progression is less than one pixel per year, therefore, it is difficult for radiologist/rheumatologist to see. Then, operating with an algorithm with pixel level accuracy to quantify JSN over a period of one year can be ineffective. JSN measured for five and ten months X-rays relative to baseline using our method are $-0.111pixel$ and $0.213pixel$ respectively.}
\label{fig:JSN_ex}
\end{figure}

Radiology plays a crucial role in diagnosis and monitoring of RA. Clinical radiologist/rheumatologist can assess the radiographic progression of RA by using the Sharp/van der Heijde scoring method (SvdH). This method relies on scoring of the radiographies by subjectively assessing JSN and bone erosion of $38$ hand or foot joints \cite{van2000read}.
JSN progression is one of the most important indicators in RA treatment, as it can directly impact medication.
Limited by the current spatial resolution of radiographic imaging, JSN progression over a period of one year can be less than one pixel, as show in Fig. \ref{fig:JSN_ex}.
This means that the pixel-level accuracy algorithm requires more time to wait for the change of joint space.
Nevertheless, this can lead to insensitive monitoring of JSN progression, and this may hinder the radiologist/rheumatologist from making a proper diagnosis in the "window of opportunity" \cite{bergstra2020earlier, kvien2006epidemiological, shimizu2019structural}.

In recent years, researchers have invested great efforts to study automatic quantification of joint space for RA diagnosis \cite{peloschek2007automatic, langs2008automatic, hirano2019development, ureten2020detection, maziarz2021deep, morita2018finger, nakatsu2020finger, duryea2000neural, angwin2001reliability, van2008automatic, bielecki2008hand, zielinski2009hand, huo2015automatic, kato2019detection, ou2019automatic, taguchi2021quantification}.
However, constrained by limited accuracy of related works, as shown in section \ref{sec:JSNquantification}, sensitively monitoring cartilage damage and bone destruction progression in the early stages of RA is a recognized challenge \cite{peloschek2007automatic, huo2015automatic}.
The major benefit of this study compared to most extant ones is an attempt to improve sensitivity and accuracy of the JSN quantification algorithm to sub-pixel level so that rheumatologist can monitor the JSN progression in RA early stages on an annual basis.

The JSN progression quantification pipline in radiographs is performed in two steps; joint position detection and joint space quantification.

\subsubsection{Related works about finger joint position detection}

The earliest studies about finger joint location detection were based on using pixel information.
Those algorithms extracted the finger midlines based on ridge detection, thus, finger joint location can be detected according to the gradient or intensity information of finger midline \cite{duryea1999automated, van2008automatic, bielecki2008hand, zielinski2009hand, huo2015automatic}. However, these method may break finger midline at the metacarpophalangeal (MCP) joint because of decrease in bone density. And may mismatch joint position for the following reasons: (i) bone overlap caused by finger bending in the vertical plane. (ii) marginal density decrease caused by ankylosis or complete luxation \cite{huo2015automatic}.

In recent years, machine learning (ML) based methods have become a very important tool to solve complex medical image processing tasks \cite{wernick2010machine}. It is widely used in image segmentation \cite{ronneberger2015u}, computer aided diagnosis \cite{asiri2019deep}, image registration \cite{haskins2020deep} and others \cite{erickson2017machine}. For finger joint detection, there are some ML-based studies utilizing key point detection for convolutional neural network (CNN) \cite{langs2008automatic, hirano2019development}, support vector machine (SVM) \cite{morita2018finger, nakatsu2020finger} and Haar-like adaptive boosting (AdaBoost) \cite{lee2015fingernet, ou2019automatic}. Although these ML-based methods improve the detection accuracy, there are still several pixels deviation in the deduced joint position. 

\subsubsection{Literature survey of joint space quantification in RA}

In the literature, detecting the upper and lower bone margins to measure the joint space width (JSW) is the most common program. They can be broadly grouped in to two groups; supervised ML-based \cite{peloschek2007automatic, langs2008automatic, hirano2019development, ureten2020detection, maziarz2021deep, morita2018finger, nakatsu2020finger} and image features based \cite{duryea2000neural, angwin2001reliability, van2008automatic, bielecki2008hand, zielinski2009hand, huo2015automatic}, such as intensity, gradient, derivative or differential.
As a representative for early RA diagnosis, Huo proposed a fully automatic JSW quantification method \cite{huo2015automatic}. Their work can be performed as follow: (i) Detect bone margin by using intensity and gradient information. (ii) Fit polynomial functions to bone margin curves. (iii) Quantify JSW according to the distance between polynomial function.
As an exception, Kato et al. proposed a JSN quantification method without margin detection \cite{kato2019detection}. This method can monitor JSN by comparing the difference of pixel intensity between the baseline and its follow-up joint windows, which is more sensitive to JSN.

For analyzing joints of RA patients, some supervised ML-based studies have been proposed like Peloschek et al \cite{peloschek2007automatic, langs2008automatic}. 
They report a JSW quantification and erosion detection method based on key point detection by using active shape models (ASMs).
Additionally, some studies have explored CNN \cite{hirano2019development, ureten2020detection, maziarz2021deep}, and SVM \cite{morita2018finger, nakatsu2020finger} to score radiographic joint destruction rather than trying to quantify the joint space.
These studies grouped finger joints into five levels according to the SvdH scoring method \cite{van2000read}.
Compared to other joint space quantification studies, scoring standard with only five or less levels limits the sensitivity and timeliness of the tool. Therefore, if it is not feasible to increase the number of scoring levels, that would make it difficult for radiologist/rheumatologist to make accurate scores for training data. In summary, these scoring-based studies can quickly determine the RA condition in the initial diagnosis but lose sensitivity in RA progression monitoring.

Scoring-based ML methods are not sensitive enough to precisely monitor RA progression. Margin detection based methods also have two main limitations: (a)
fundamentally detecting the upper bone margin accurately is a perceived challenge, which is known to be affected by false edges \cite{van2008automatic, angwin2001reliability, huo2015automatic}.
(b) Margin detection based studies \cite{peloschek2007automatic, langs2008automatic, hirano2019development, ureten2020detection, maziarz2021deep, morita2018finger, nakatsu2020finger, duryea2000neural, angwin2001reliability, van2008automatic, bielecki2008hand, zielinski2009hand, huo2015automatic, kato2019detection} can best achieve a pixel-level accuracy only (please see Section \ref{sec:JSNquantification} for more details). This can impact the timeliness of clinical decision making in RA. Although, intensity difference based JSN quantification method \cite{kato2019detection} has better sensitivity than others, but it is susceptible to imaging environments.

\begin{figure*}[!t]
\centering\includegraphics[width=\textwidth, trim=0 30 0 0, clip]{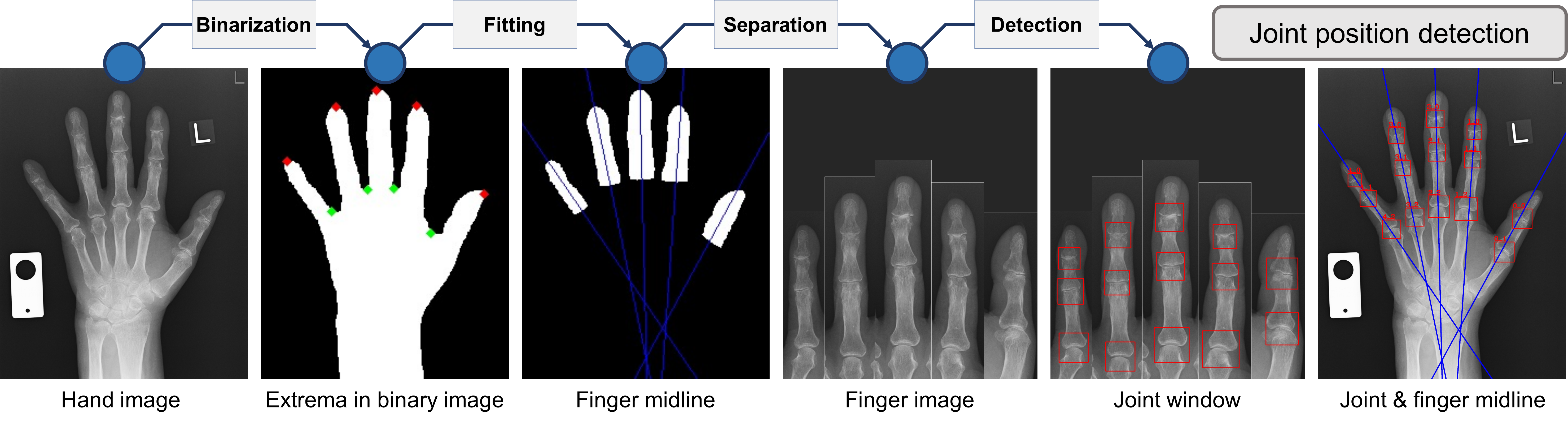}
\caption{Schematic overview of joint position detection. The approximate areas of fingers are obtained according to each pair of local maxima (red points) and local minima (green points) in binary image. Then, the finger midlines (blue lines) are calculated by fitting to each area. Finally, an AdaBoost based joint classifier is used to detect the joint positions (red windows).}
\label{fig:JPD}
\end{figure*}

\subsubsection{Our contributions}
Original contribution of our work can be summarized as follows:
(i) Describe a detection method for finger midline and joint position. 
(ii) Propose an image segmentation algorithm to
segment joint images. 
(iii) Present an improved phase-only correlation method named partial image phase-only correlation (PIPOC) to measure JSN progression in the early RA stage. (iv) Automate the features listed in (i-iii). Using our method the JSN progression can be measured from a group of input sequential radiographs. 
(v) The proposed work can achieve sub-pixel accuracy on JSN progression measurement.

This rest of the paper is organized as follows: Section II reports a fully automatic method for the localization of the joint position, and also describes the  quantification of the JSN progression in detail. In Section III, we describe our experiments; including phantom imaging, and clinical imaging. Section IV, presents the joint position detection results using clinical data and the evaluation results for both phantom imaging and clinical imaging for JSN progression quantification. Section V presents a detailed discussion with concluding remarks.

\section{METHODOLOGY}
\label{sec:method}

The main objective of the proposed JSN quantification algorithm is to improve sensitivity, accuracy and robustness so that radiologist/rheumatologist can closely monitor the JSN progression in RA at an early stage.
Ideally, the automatic quantification of JSN progression in finger joints radiographs is performed in three steps.
(i) Detect and calibrate joint positions. (ii) Segment upper and lower bones of joints based on gradient information. (iii) Measure the location differences of upper and lower bones between baseline and follow-up hand radiographs respectively by using PIPOC, thus resulting in JSN quantification.  

\subsection{Joint position detection and calibration}

As shown in Fig. \ref{fig:JPD}, the pipeline of joint position detection and calibration can be briefly explained as follows: (i) Obtain the approximate estimates of the finger midlines in binary image. (ii) Detect joint positions by using a ML-based joint classifier. (iii) Calibrate the relative position deviations in joint windows.

\subsubsection{Finger midline detection}

Finger position estimation can significantly reduce the potential area, thus reduces the calculation of joint detection. The scheme of finger midline detection is shown in Fig. \ref{fig:JPD}, the approximate area and angle of fingers are estimated using the binary image obtained from a hand radiograph.

Given a radiograph, we binarize the X-ray using Otsu's method \cite{otsu1979threshold}, and smooth its margin by using morphological opening and closing \cite{haralick1987image}.
We obtain the local maxima (red point) and local minima (green point) of hand margins as shown in Fig. \ref{fig:JPD}.
From our experiments we found that using polygonal approximation can significantly improve robustness when searching for extrema, that one could obtain using pure margin \cite{sklansky1980fast}. 
Next, the approximate area of fingers are obtained according to each pair of local maxima and minima, which can calculate the midline of each finger by fitting to each finger area based on least squares method \cite{york1968least}.

From our experiments and analysis we found that reducing the width and height of the binary hand images to one-fifth does not significantly effect the accuracy of the finger midline detection, and this results in accelerating the detection process ($17.7\times$ faster).

\begin{figure}[!t]
\centering\includegraphics[width=\linewidth, trim=0 30 0 0, clip]{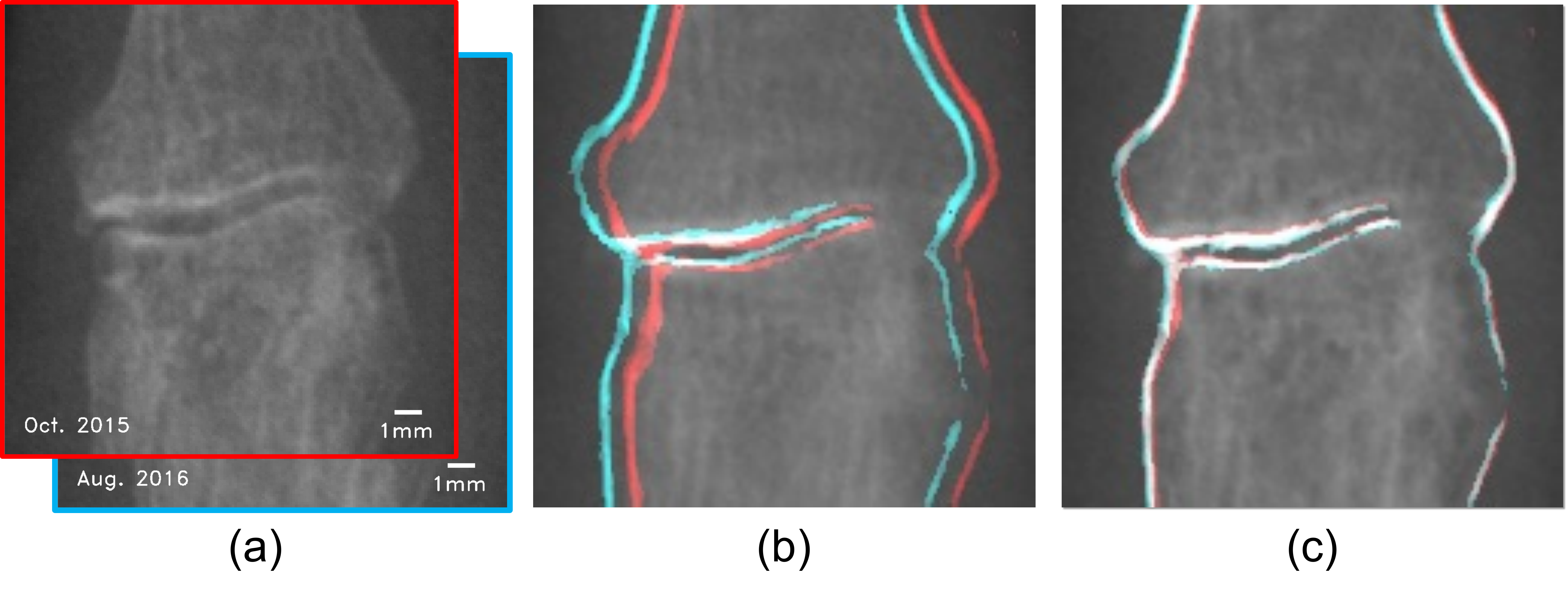}
\caption{Results of joint position calibration: (a) A proximal interphalangeal joint (PIP) of little finger in baseline radiograph (red border) and follow-up radiograph (cyan border). (b) The margin of PIP radiograph in (a) before position calibration (red: baseline radiograph, cyan: follow-up radiograph, white: overlap). (c) The margin information after position calibration.}
\label{fig:calibration}
\end{figure}

\begin{figure*}[!t]
\centering\includegraphics[width=\textwidth, trim=0 30 0 0, clip]{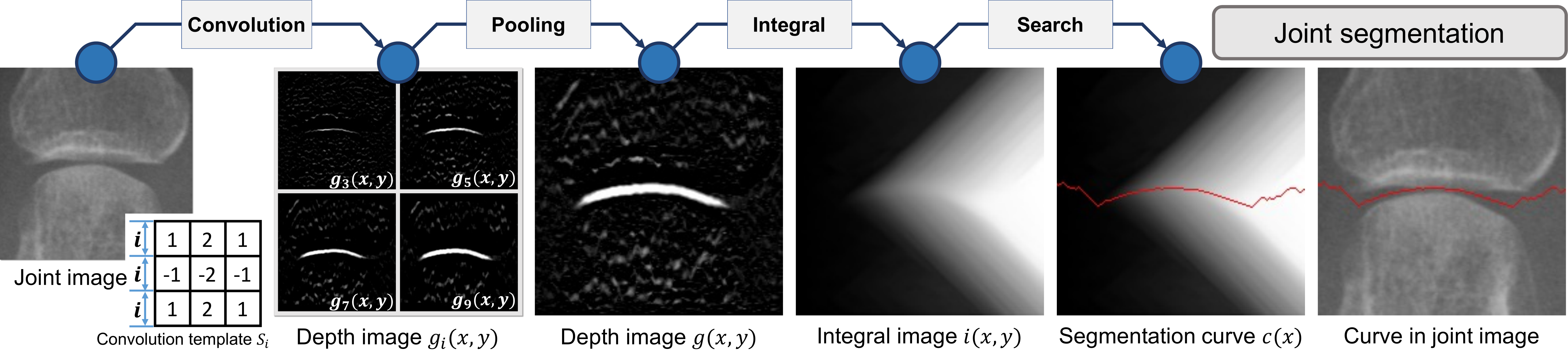}
\caption{Overview of joint segmentation based on gradient information. Gully depth map $g(x, y)$ is calculated to define depth feather.
Independent edge filtering on the upper and lower side determines the pixel depth. Height-adjustable convolution template $S_i$ ensures that a given range of gully can be detected. The integral image $i(x, y)$ is calculated to search the segmentation curve $c(x)$ with the maximum depth-sum.}
\label{fig:segmentation}
\end{figure*}

\subsubsection{Joint position detection}
As shown in Fig. \ref{fig:JPD}, the position, angle, and size of each finger image is obtained according to the finger midline and its area in the binary image. 
Then, the joint windows are detected in finger images with a joint classifier which is trained by using haar-like feature based AdaBoost \cite{viola2001rapid, lienhart2002extended, lienhart2003empirical}. 

\subsubsection{Joint position calibration}
We propose a low computational solution based on FIPOC to calibrate the joint position, a detailed discussion of FIPOC implementation is presented in section \ref{sec:poc}.

The bone surface texture is frequently changing and varies when the bone is growing, and these irregular variation will cause phase dispersion in phase difference spectrum when using phase-only correlation method.
Our experiments show that edge preserving filter can significantly reduce mismatch error of phase-only correlation, which can suppress bone surface texture information while preserving the bone margin.
In our work, we preprocess the image using a median filter to calibrate position deviation \cite{huang1979fast}.

As show in Fig. \ref{fig:calibration} the joint position calibration which relies on FIPOC cannot reduce the deviation with ground truth. It can limit relative position deviation between baseline and follow-up joint windows within one pixel.

\subsection{Joint segmentation}

In this stage, the upper and lower bones are segmented from the joint image, based on gradient information, so that the displacements of the upper and lower bones can be mearsured separately.

\subsubsection{Depth image} 

The depth map is used to gauge the depth of each pixel within a given range of width. Only the vertical depth is detected in this work, because all joint images are arranged vertically. Nevertheless, the depth of any direction can be detected with a customized convolution template $S_i$. The detailed explanation is shown in Fig. \ref{fig:segmentation}.

In order to detect depth within a range of width, a height-adjustable convolution template $S_i$ is used ($i$ is odd) to calculate the depth of $i$ pixels gully width. Consider an image with $M$ pixel width and $N$ pixel height, the convolutions of $S_{ia}$ and $S_{ib}$ are as follows:
\begin{equation}
\begin{split}
& g_{ia}(x,y)=\sum_{k=-1}^{1}\sum_{l=-(i-1)/2}^{(3i-1)/2}s_{i}(k,l)f(x+k,y+l)\\
& g_{ib}(x,y)=\sum_{k=-1}^{1}\sum_{l=-(3i-1)/2}^{(i-1)/2}s_{i}(k,l)f(x+k,y+l)
\end{split}
\end{equation}

$g_{ia}(x,y)$ is the grayscale difference between the current line with the upper line, and $g_{ib}(x,y)$ is the grayscale difference between the current line with the lower line. The smaller value of $g_{ia}(x,y)$ and $g_{ib}(x,y)$ is defined as the depth, as shown in Eq. \ref{eq:g_i}.
\begin{equation}
g_i(x,y)=\min(g_{ia}(x,y),g_{ib}(x,y))
\label{eq:g_i}
\end{equation}

In this work, the range of width $[i_{min}, i_{max}]$ is defined as one to nine pixels when the spatial resolution is $0.175mm/pixel$. Next, we perform maximum pooling on depth map $g_i$, as show in Eq. \ref{eq:pooling}.
\begin{equation}
g(x,y)=\max(g_{i_{min}}(x,y),\cdots, g_{i_{max}}(x,y))
\label{eq:pooling}
\end{equation}

As shown in Fig. \ref{fig:segmentation}, the depth map $g(x,y)$ corresponds to the depth information by using gradient information.

\subsubsection{Integral image}

The integral image $i(x, y)$ is an intermediate matrix, which is used to find the segmentation curve with the maximum depth-sum. It can be expressed as the local maximum in the left column plus depth map $g(x,y)$, as shown in Eq. \ref{eq:integral}.
\begin{small}
\begin{equation}
i(x,\!y)\!\!=\!\!\begin{cases}
\!g(x,\!y) & \!\!\!x\!\!=\!\!0 \\
\!\max(i(\!x\!\!-\!\!1,\!y\!\!-\!\!1)\!,\!i(x\!\!-\!\!1,\!y)\!,\!i(x\!\!-\!\!1,\!y\!\!+\!\!1)\!)\!\!+\!\!g(x,\!y) & \!\!\!x\!\!>\!\!0\\
\end{cases}
\label{eq:integral}
\end{equation}
\end{small}

\subsubsection{Segmentation curve}

The segmentation curve with the maximum depth-sum can be determined from integral image $i(x, y)$ as follows. First, determine the maximum value of the rightmost column in $i(x, y)$ as the end point of the segmentation curve. Then, select the maximum of the three adjacent pixels in the left column in $i(x, y)$ as the next point of the segmentation curve until arriving at the leftmost column. The segmentation curve $c(x)$ is defined as Eq. \ref{eq:segmentationCurve}. The $\mathop{\arg\max}_{y} i(x, y)$ indicates the index of the maximum value on the $y$ axis for a given $x$ value in a given $y$ range.
\begin{equation}
c(x)\!\!=\!\!\mathop{\arg\max}_{y} i(x, \!y)\!\!\begin{cases}
\!y\!\!\in\!\![0, N\!\!-\!\!1] & \!\!\!x\!\!=\!\!M\!\!-\!\!1 \\
\!y\!\!\in\!\![c(\!x\!\!+\!\!1)\!\!-\!\!1, c(\!x\!\!+\!\!1)\!\!+\!\!1] & \!\!\!x\!\!<\!\!M\!\!-\!\!1 \\
\end{cases}
\label{eq:segmentationCurve}
\end{equation}

The binary matrix of upper bone area $s_0(x, y)$ and lower bone area $s_1(x, y)$ can be expressed as Eq. \ref{eq:segmentationArea}, according to the segmentation curve $c(x)$.
\begin{equation}
\begin{split}
& s_0(x, y)=\begin{cases}
1 & y<c(x)\\
0 & \text{otherwise}
\end{cases}
\\
& s_1(x, y)=\begin{cases}
1 & y>c(x)\\
0 & \text{otherwise}
\end{cases}
\end{split}
\label{eq:segmentationArea}
\end{equation}

An example of finger joint segmentation is shown in Fig. \ref{fig:segmentation}.

\begin{figure*}[!t]
\centering\includegraphics[width=\textwidth]{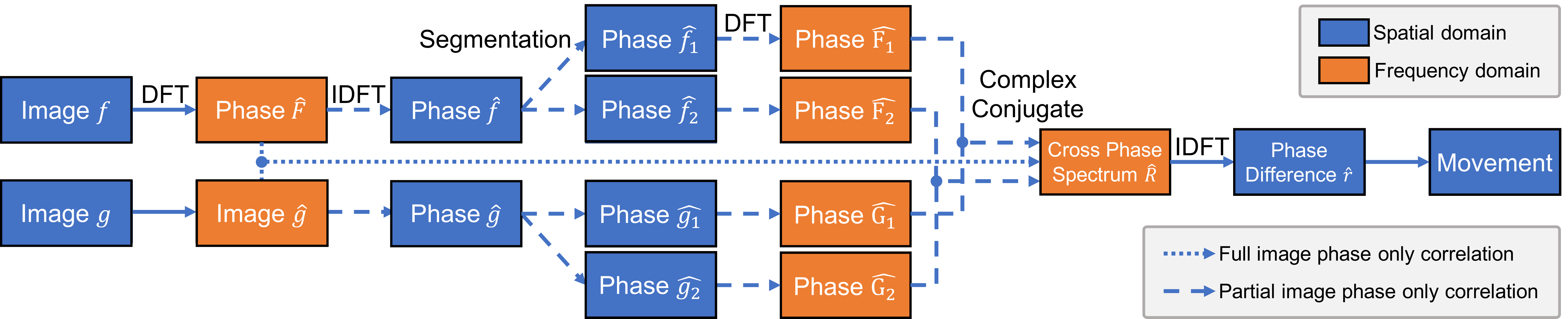}
\caption{A flowchart describing the sequence of operations for implementing FIPOC and PIPOC algorithms.}
\label{fig:teaser}
\end{figure*}

\subsection{JSN quantification by partial image phase-only correlation}
\label{sec:poc}

The basic concept of JSN quantification by FIPOC or PIPOC can be described using the flowchart in Fig \ref{fig:teaser}. Consider two images, $f(x, y)$ and $g(x, y)$, which are divided into $k$ areas. Consider an area $i$, let $\alpha_i$ and $\beta_i$ represent sub-pixel displacement from $f(x, y)$ to $g(x, y)$ in $x$ and $y$ directions respectively, and a binary matrix $s_i(x, y)$ that includes segmentation information. So, $g(x, y)$ can be represented as Eq. \ref{eq:deviation}. 
\begin{equation}
g(x, y)=\sum\nolimits_{i=0}^{k}f(x-\alpha _i, y-\beta _i)*s_i(x, y)
\label{eq:deviation}
\end{equation}

A 2D Hanning window function is applied to input images $f(x, y)$ and $g(x, y)$ to reduce the effect of discontinuity at image border \cite{takita2003high}. The Hanning window $w(x,y)$ can be defined as:
\begin{equation}
w(x, y)=\frac{1+cos(\frac{\pi x}{M})}{2}\frac{1+cos(\frac{\pi y}{N})}{2}
\end{equation}

Let $F(u, v)$ and $G(u, v)$ denote the 2D Discrete Fourier Transforms (DFT) of the two images. Considering the properties of DFT $\mathcal{F}$, $F(u, v)$ and $G(u, v)$ can be expressed as Eq. \ref{eq:DFT}.
\begin{equation}
F(u,\!v)\!\!=\!\!\mathcal{F}(f(x,\!y)w(x,\!y))\quad G(u,\!v)\!\!=\!\!\mathcal{F}(g(x,\!y)w(x,\!y))
\label{eq:DFT}
\end{equation}

Next, extract the phase component of $F(u, v)$ and $G(u, v)$, the functions are divided by the amplitude, as follows:
\begin{equation}
\hat{F}(u, v)=\frac{F(u, v)}{|F(u, v)|}\quad \hat{G}(u, v)=\frac{G(u, v)}{|G(u, v)|}
\end{equation}

FIPOC will calculate the phase difference spectrum $\hat{r}(u, v)$ between $\hat{F}(u, v)$ and $\hat{G}(u, v)$ immediately (the dotted line in Fig. \ref{fig:teaser}). But when the displacement of each area is different, there will be several dirac delta functions in phase difference spectrum, as show in Eq. \ref{eq:FIPOC}. 
\begin{equation}
\hat{r}(u, v)=\sum\nolimits_{i=0}^{k}p_i\delta(\alpha_i, \beta_i)
\label{eq:FIPOC}
\end{equation}

Different from FIPOC, PIPOC segments the phase spectrum in spatial domain. Next, the phase spectrum $\hat{f}(x, y)$ of image $f(x, y)$ and the phase spectrum $\hat{g}(x, y)$ of image $g(x, y)$ in spatial domain are obtained by Inverse Discrete Fourier Transform (IDFT) $\mathcal{F}^{-1}$.
\begin{equation}
\hat{f}(x, y)=\mathcal{F}^{-1}(\hat{F}(u, v)) \quad \hat{g}(x, y)=\mathcal{F}^{-1}(\hat{G}(u, v))
\end{equation}

Segmenting area $i$ by using segmentation matrix $s_i(x, y)$.
\begin{equation}
\begin{split}
\hat{f_i}(x, y)\!=\!\hat{f}(x, y)\!*\!s_i(x, y) \quad \hat{g_i}(x, y)\!=\!\hat{g}(x, y)\!*\!s_i(x, y)
\end{split}
\label{eq:phase}
\end{equation}

Subsequently combining DFT $\mathcal{F}$ and Eq. \ref{eq:phase} to develop the phase spectrum of area $i$ in frequency domain.
\begin{equation}
\begin{split}
\hat{F_i}(u, v)=\mathcal{F}(\hat{f_i}(x, y)) \quad
\hat{G_i}(u, v)=\mathcal{F}(\hat{g_i}(x, y))
\end{split}
\end{equation}

The normalized cross phase spectrum $\hat{R_i}(u, v)$ of area $i$ between $F(u, v)$ and $G(u, v)$ can be obtained respectively as given in Eq. \ref{eq:NCPS}. Here, $\overline{G_i(u, v)}$ in Eq. \ref{eq:NCPS} denotes the complex conjugate of ${G_i(u, v)}$.
\begin{equation}
\hat{R_i}(u, v)=\frac{\hat{F_i}(u, v)\overline{\hat{G_i}(u, v)}}{|\hat{F_i}(u, v)\overline{\hat{G_i}(u, v)}|}
\label{eq:NCPS}
\end{equation}

Next, the phase difference spectrums $\hat{r_i}(x, y)$ of area $i$ between the two images are obtained by IDFT $\mathcal{F}^{-1}$. The location of the dirac delta function $\delta$ represents the displacement between two images.
\begin{align}
\hat{r_i}(x, y)
\nonumber & =\mathcal{F}^{-1}(\hat{R_i}(u, v))\\
& =\delta(\alpha _i, \beta _i)
\end{align}

In case of Fourier Transform, the location of the peak of dirac delta function $\delta$ in the phase difference spectrum $\hat{r_i}(x, y)$ can be determined according to the maximum peak.
\begin{equation}
(\alpha _i{'}, \beta _i{'}) = \mathop{\arg\max}_{(x, y)}\hat{r_i}(x, y)
\label{eq:PIPOC}
\end{equation}

Consider the DFT, the least-square fitting method employed to estimate displacement $(\alpha _i, \beta _i)$ around the maximum peak $(\alpha _i{'}, \beta _i{'})$. Since the $\delta$ function has a very sharp peak, limited number of data points $5\times 5$ are used to fit $\delta$ function \cite{takita2003high} in this work. Thus, the $\text{JSN}_{fg}$ between image $f(x, y)$ and image $g(x, y)$ can be quantified according to the displacement difference between upper bone area $s_0(x, y)$ and lower bone area $s_1(x, y)$, as show below.
\begin{equation}
\text{JSN}_{fg} = \beta_0-\beta_1
\label{eq:JSN}
\end{equation}

\begin{figure}[!t]
\centering\includegraphics[width=\linewidth, trim=10 30 0 10, clip]{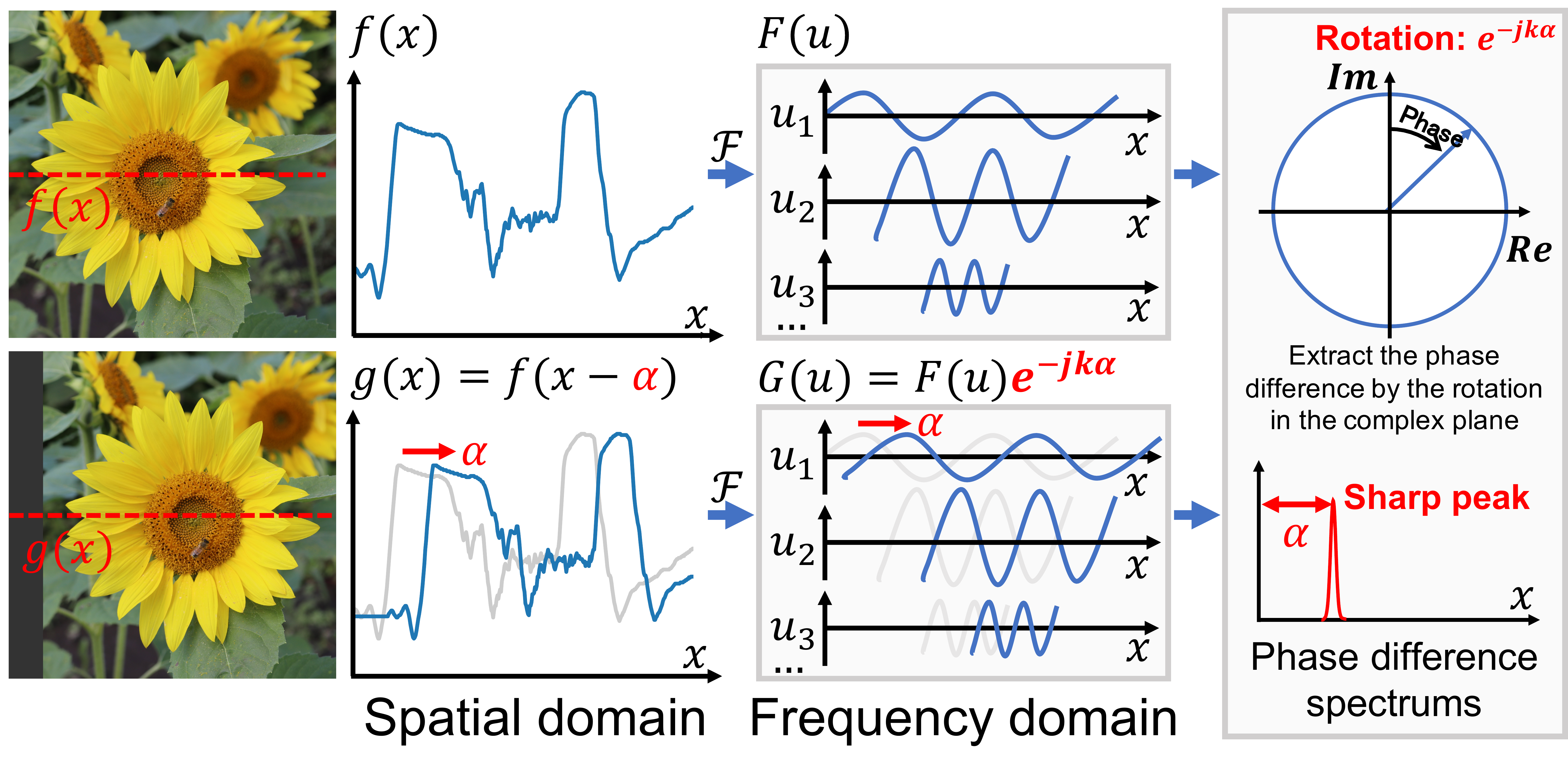}
\caption{Principles of FIPOC. Consider two signals $f(x)$ and $g(x)$ with $\alpha$ displacement. Since each wave has the same phase difference, the displacement can be measured by the location of dirac delta function in the phase difference spectrum.} 
\label{fig:POC}
\end{figure}

\begin{figure*}[!t]
\centering\includegraphics[width=\textwidth, trim=20 20 0 0, clip]{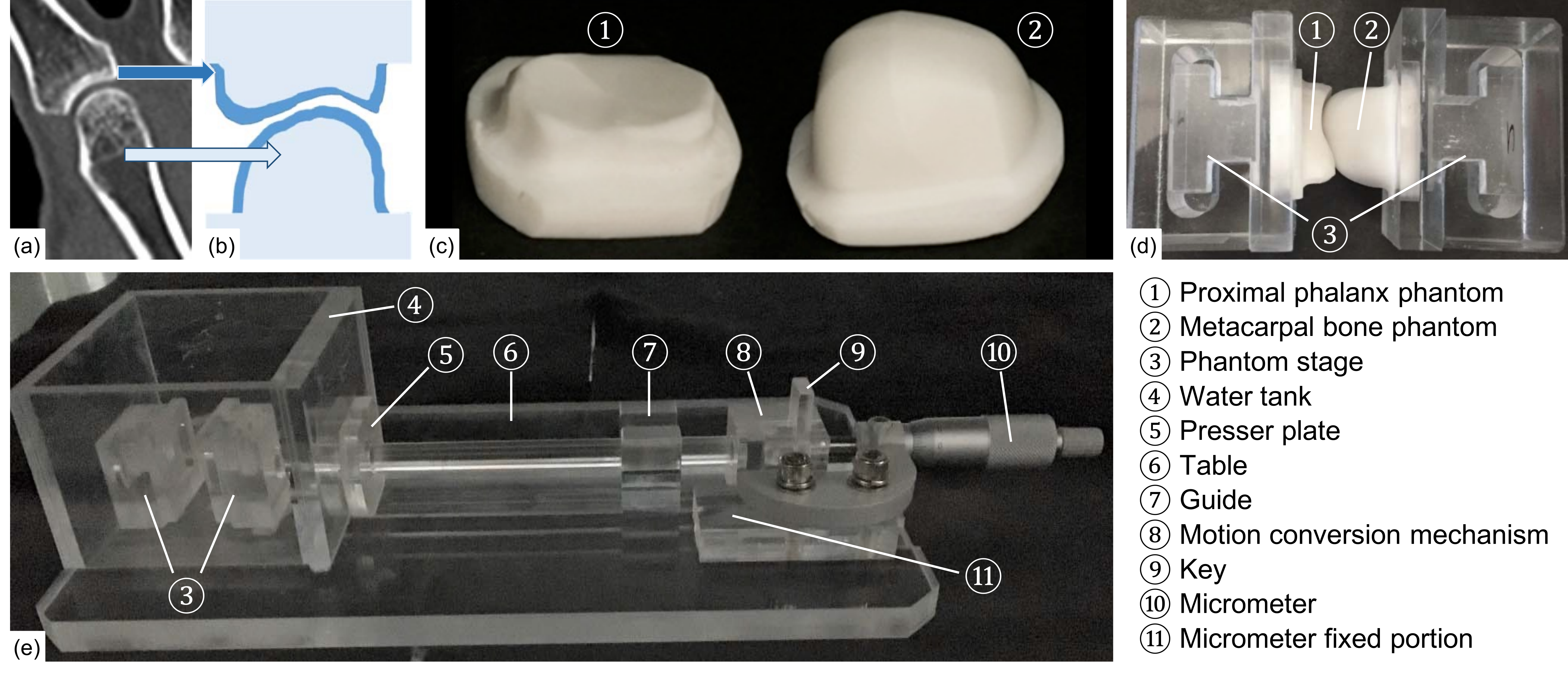}
\caption{A MCP joint-shaped two-layer phantom design and phantom imaging environment. (a) A MCP joint in radiographic clinical imaging. (b) A diagram of two-layer structure bone (dark blue: Bone cortex, light blue: Cancellous bone). (c) A set of MCP joint-shaped two-layer phantom. (d) The phantom joint connect with the attaching portion. (e) The phantom imaging environment.}
\label{fig:device}
\end{figure*}

\subsection{PIPOC and FIPOC}
\label{poc}

Earlier we had proposed a JSN quantification method \cite{ou2019automatic, taguchi2021quantification}, which is based on full image phase-only correlation (FIPOC) \cite{reddy1996fft, stone2001fast, foroosh2002extension, takita2003high}. 
FIPOC is a well-known method for image registration, it can estimate the relative displacement between two images and it relies on the frequency domain analysis.
Figure \ref{fig:POC} shows a one dimension diagram of FIPOC, it has sub-pixel level accuracy and error range within $0.01 pixel$ when measured on a set of $200 pixel\times200 pixel$ images \cite{takita2003high}, and the accuracy improves with higher image resolution.

FIPOC can calculate displacement between two images by measuring the phase difference in frequency domain. However, it has a limitation when there are multiple areas with different displacements. Each local area would have a corresponding independent dirac delta function in phase difference spectrum. 
The precise position of each dirac delta function can be obtained if and only if the displacement differences between multiple areas are large enough (about $3 pixel$ \cite{shimada2018sparse}). Otherwise, dirac delta functions in phase difference sepctrum will be affected and even overlap.

\begin{figure}[!t]
\centering\includegraphics[width=\linewidth]{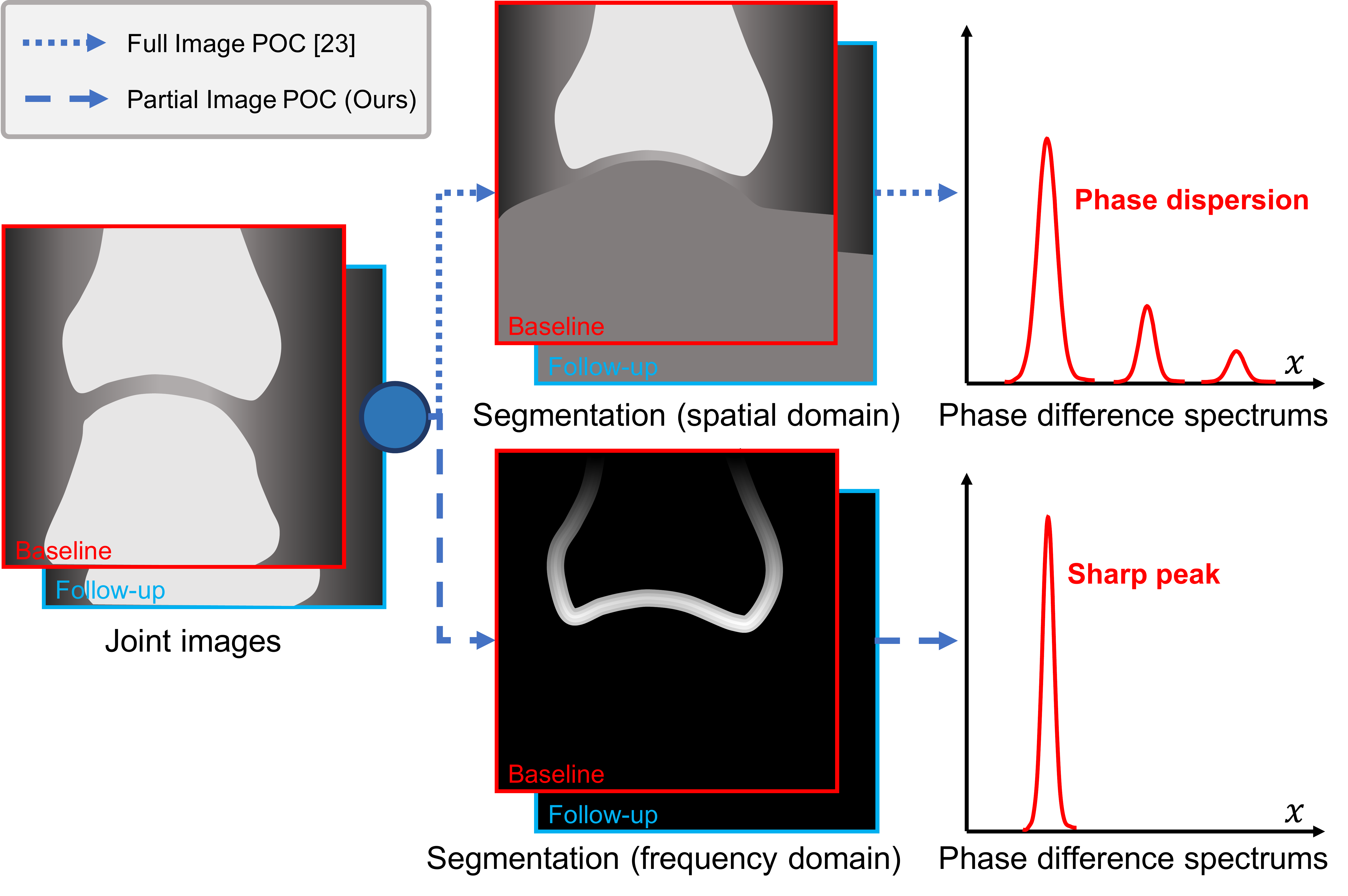}
\caption{Difference between FIPOC and PIPOC for JSN measurement. FIPOC-based JSN measurement requires the segmentation of upper and lower bone, and in-painting algorithm is used to fill vacant space, this could lead to phase dispersion as shown above. In this work (PIPOC) we obtain a sharper delta function by quantifying the displacement by segmenting joint image in the phase spectrum of the frequency domain.}
\label{fig:PIPOC_FIPOC}
\end{figure} 

In our conference paper \cite{ou2019automatic}, we segmented joint images in spatial domain and fill vacant space by using image in-painting algorithm \cite{telea2004image} to solve the overlapping issue. Thus, the displacements of upper and lower bones could be quantified from independent bone images based on FIPOC respectively. But some non-existent phase features may be generated which could increase error and even mismatches, as shown in Fig. \ref{fig:PIPOC_FIPOC}. To avoid the impact from in-painting algorithm, the partial image phase-only correlation (PIPOC) is proposed in this work. The basic idea of PIPOC is to determine the location of each dirac delta functions by segmenting the phase spectrum. Compared to FIPOC combined image in-painting, PIPOC can further improve the accuracy and robustness of JSN quantification, by eliminating the impact of image in-painting algorithm.

\section{Materials}
\label{sec:material}

Imaging phantom based experiments were studied to evaluate our algorithm's performance. From our experiments we observed that the manually labeled JSW has pixel level mean error, and it is discussed in Section \ref{sec:phantomStudy}. We prepared phantom images with ground truth to evaluate our algorithm in terms of absolute error. Additionally, our algorithm's quality performance was also evaluated using clinical data.

\subsection{Phantom imaging}

\subsubsection{Phantom design}

A phalanx-shaped phantom was produced using vacuum-sintered bodies of a novel apatite called Titanium medical apatite (TMA) \cite{tamura2013vacuum}. The chemical formula of TMA is $Ca_{10}(PO_4)_6\cdot TiO_2\cdot (OH)_2\cdot nH_2O$. TMA powder was kneaded with distilled water, and solid cylinders of compacted TMA were formed by compression molding at $10 MPa$. TMA was vacuum sintered using a resistance furnace at about $10^{-3} Pa$.

Using TMA to design imaging phantom has the following advantages: (1) The CT value of phantom in radiographs can be easily modified by changing the ratio of TMA and adhesive. 
(2) TMA bodies are easy to process and model with a 3D-modeling machine or a lathe. (3) TMA vacuum-sintered bodies has a density of approximately $2300 kg/m^3$ (corresponding to that of a compact bone or a tooth).

The phantom used in our experiment is a two-layer TMA vacuum-sintered body to simulate the X-ray absorption coefficient (CT value) of bone cortex and cancellous bone. The diagram of the two-layered bone is shown in Fig. \ref{fig:device} (a) and (b). 
The phantom mimics MCP joint, proximal phalanx, and the metacarpal bone. The assembled phantoms are shown in Fig. \ref{fig:device} (c). The important properties of our phantom bones are given in Table \ref{tab:phantomParaments} \cite{kato2019detection, tamura2013vacuum}.

\begin{table}[!t]
\renewcommand{\arraystretch}{1}
\caption{Phantom design preparation}
\centering
\setlength{\tabcolsep}{2.165mm}{
\begin{tabular*}{\linewidth}{p{3.5cm}<{\centering}|p{2cm}<{\centering}p{2cm}<{\centering}}
\bottomrule
~ & Bone cortex & Cancellous bone \\ \hline
TMA : adhesive & 1:1.2 & 1:5 \\
Particle Size ($\mu$m) & 107 $\sim$ 250 & 107 $\sim$ 250 \\
Temperature (K) & 1370 & 1370 \\ \toprule
\end{tabular*}
}
\label{tab:phantomParaments}
\end{table} 

\begin{table}[!t]
\renewcommand{\arraystretch}{1}
\caption{X-ray imaging configuration parameters}
\centering
\setlength{\tabcolsep}{2.165mm}{
\begin{tabular*}{\linewidth}{p{3.5cm}<{\centering}|p{2cm}<{\centering}p{2cm}<{\centering}}
\bottomrule
	~ & Phantom & Clinical \\ \hline
	Tube voltage (kV) & 50 & 42 \\
	Tube current (mA) & 100 & 100 \\
	Exposure time (mSec) & 20 & 20 \\
	Source to image (cm) & 100 & 100 \\ \toprule
\end{tabular*}
}
\label{tab:imagingParaments}
\end{table}

\subsubsection{Imaging environment}

The phantom joint was mounted on to the stage as shown in Fig. \ref{fig:device} (d). The phantom stage was connected to a micrometer, and thereby the JSW of phantom could be easily adjusted using the micrometer controls. The JSW range is up to $13mm$, and has a minimum scale of $0.01mm$. There is substantial evidence that JSW has a close relationship with age and sex in healthy populations \cite{kvien2006epidemiological, pfeil2007computer}. In addition, RA is more frequent in females who are between $30$ and $50$ years of age, and their JSW is around $1.70mm$ \cite{kvien2006epidemiological, pfeil2007computer}. In our work, the JSW standard of phantom was set as $1.70mm$. Following two sets of phantom images with different specifications were provided. (i) JSW range: $1.20mm$ - $2.20mm$, increment step size: $0.10mm$ (ii) JSW range: $1.65mm$ - $1.75mm$, increment step size: $0.01mm$.

Considering that in clinical data, the X-ray beam can be attenuated by the tissue, these attenuations are displayed as noise in the radiography. In related phantom studies, water is usually used to simulate the noise generated by the beam attenuation in the tissue \cite{brooks1976statistical, chesler1977noise}. In our experiment, the phalanx-shaped phantom was mounted on the stage as shown in Fig. \ref{fig:device} (d), and placed in a tank. We can image the phantom with low noise when the tank is filled with air, or filled with distilled water which has an X-ray absorbing properties similar to normal tissue. Our experimental phantom imaging setup is shown in Fig. \ref{fig:device} (e). The X-ray imaging equipment used in our experiment is: digital radiography (\textit{FUJIFILM DR CALNEO Smart C47}, \textit{Fujiflm} Corporation, Tokyo, Japan), and its parameters are given in Table \ref{tab:imagingParaments}. The spatial resolution used in this phantom study is $0.15mm/pixel$.

\subsection{Clinical dataset}

\subsubsection{Study population}

For clinical assessment, we prepared dataset from \textit{Sagawa Akira Rheumatology Clinic} (Sapporo, Japan), \textit{Sapporo City General Hospital} (Sapporo, Japan) and \textit{Hokkaido Medical Center for Rheumatic Diseases} (Sapporo, Japan). This dataset contains $1120$ hand posteroanterior projection (PA) radiographs from patients in the early RA stage. All images were used in the joint position detection experiments. Considering that several images were required to evaluate our work when calculating standard deviation. Thus, images of patients who were radiographed at least three times were retained, which contains 549 hand PA radiographs of $77$ RA patients out of which $88.0\%$ are female. Detailed patients information are summarized in Table \ref{tab:clinicalInformation} (please note, the gender and age information of a small number of patients were not included upon patient request).

This study was conducted in accordance with the guidelines of the Declaration of Helsinki and approved by the Ethics Committee of the Faculty of Health Sciences, Hokkaido University (approval number: $19-46$).

\begin{table}[!t]
\renewcommand{\arraystretch}{1}
\caption{Patient information in the clinical dataset}
\centering
\begin{threeparttable}
\setlength{\tabcolsep}{2.165mm}{
\begin{tabular*}{\linewidth}{p{3.5cm}<{\centering}|p{2cm}<{\centering}p{2cm}<{\centering}}
\bottomrule
	~ & Mean $\pm$ SD & Range \\ \hline
	Age at enrollment (year) & 55.83 $\pm$ 13.86 & 20.68 $\sim$ 88.00 \\
	Number of Photography$^*$ & 4.30 $\pm$ 2.54 & 3 $\sim$ 17 \\
	Treatment Time (year) & 4.01 $\pm$ 3.43 & 0.88 $\sim$ 12.10 \\ \toprule
\end{tabular*}
}
\begin{tablenotes}
\item[*] Patients did two-handed or one-handed radiographic imaging.
\end{tablenotes}
\label{tab:clinicalInformation}
\end{threeparttable}
\end{table}

\subsubsection{Imaging environment}

The radiographic imaging device used in our clinical study is \textit{DR-155HS2-5} from \textit{Hitachi} Corporation, with $1.5mm$ X-ray aluminum filter thickness. The centering point of the X-ray beam was the MCP joint of the middle finger. Digital imaging and communications in medicine (DICOM) standard was used in managing our radiographic dataset, and the image resolution is $2010\times1490$, and a $0.175\times0.175mm$ pixel size at $12$ bit depth. For detailed imaging parameter descriptions, please refer to Table \ref{tab:imagingParaments}.

\begin{figure}[!t]
\centering\includegraphics[width=0.75\linewidth]{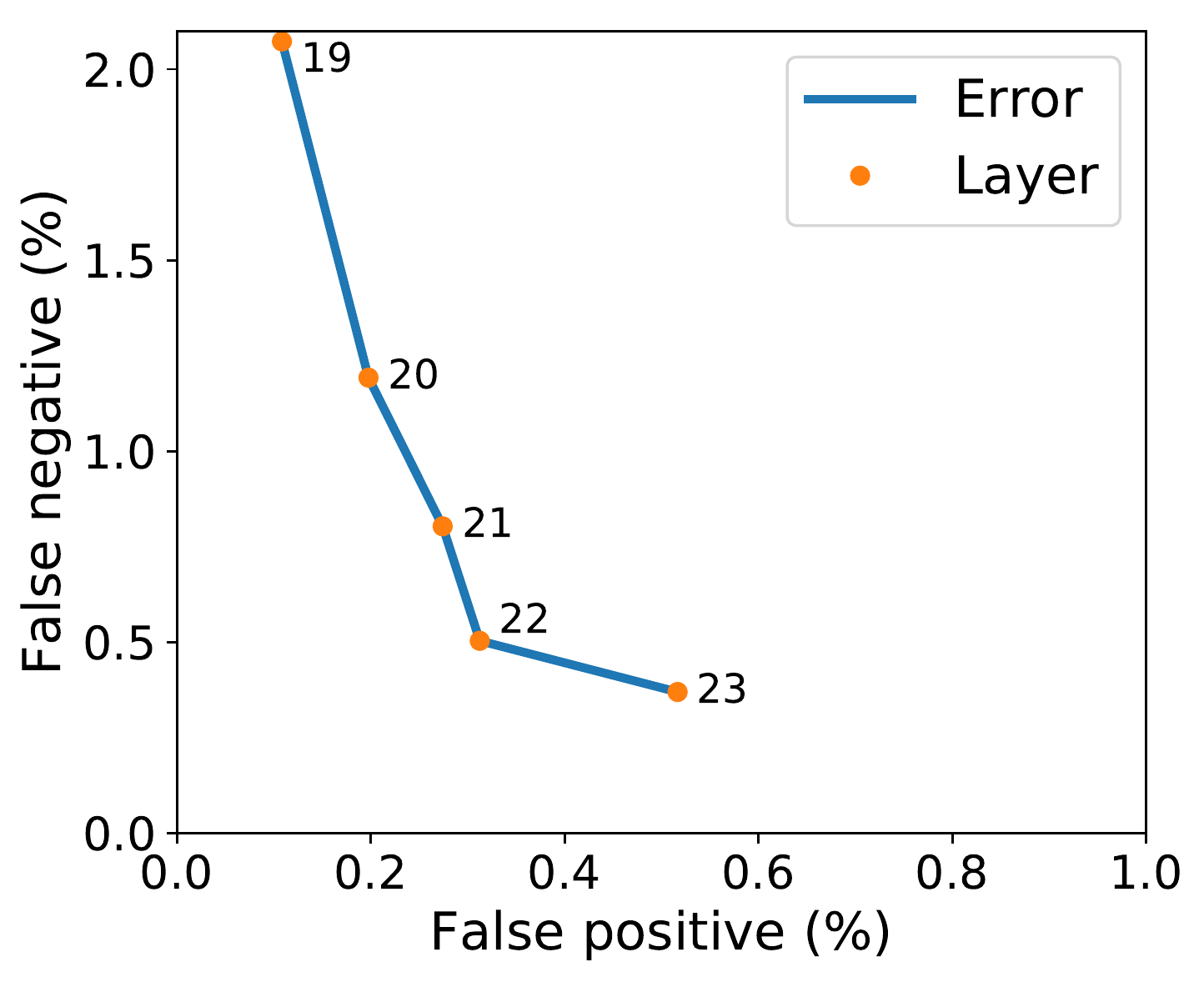}
\caption{The variation in false positive and false negative ratios with the increase of the cascade layers.}
\label{fig:detectionResult}
\end{figure}

\begin{figure*}[!t]
\centering\includegraphics[width=\textwidth, trim=0 40 0 0, clip]{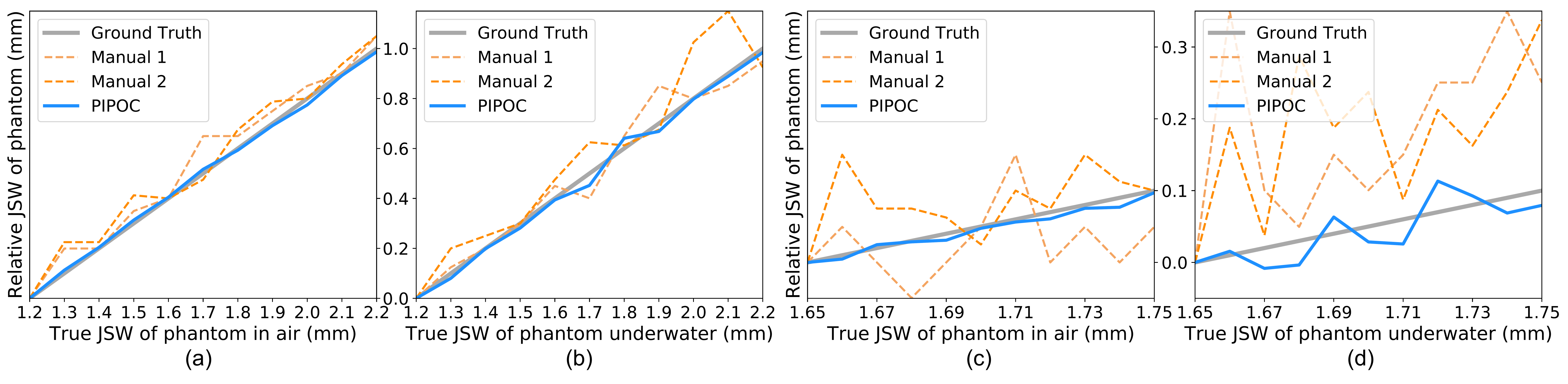}
\caption{The measurement result of PIPOC and manual when using phantom images. Blue lines are the relative JSW of each image to the first image obtained by PIPOC. Orange dot lines are the difference of manually measured JSW between every image and the first image. The true JSW of phantom is from $1.20mm$ to $2.20mm$ at increments of $0.10mm$ in sub-figure (a) and (b). And it is from $1.65mm$ to $1.75mm$ at increments of $0.01mm$ in sub-figure (c) and (d). The phantom of sub-figure (a) and (c) is placed in air. And the phantom of sub-figure (b) and (d) is placed in distilled water.}
\label{fig:phantomResult}
\end{figure*}

\begin{table*}[!t]
\renewcommand{\arraystretch}{1}
\caption{The performance analysis in millimeter for PIPOC and manual measurement when using phantom images}
\setlength{\tabcolsep}{1.262mm}{
\begin{tabular*}{\linewidth}{l|ccc|ccc|ccc|ccc}
\bottomrule
	& \multicolumn{6}{c|}{Mean Error} & \multicolumn{6}{c}{Root-Mean-Square Deviation} \\ \cline{2-13}
	& \multicolumn{3}{c|}{Air} & \multicolumn{3}{c|}{Water} & \multicolumn{3}{c|}{Air} & \multicolumn{3}{c}{Water} \\ \cline{2-13}
	& Fig.\ref{fig:phantomResult} (a) & Fig.\ref{fig:phantomResult} (c) & Average & Fig.\ref{fig:phantomResult} (b) & Fig.\ref{fig:phantomResult} (d) & Average & Fig.\ref{fig:phantomResult} (a) & Fig.\ref{fig:phantomResult} (c) & Average & Fig.\ref{fig:phantomResult} (b) & Fig.\ref{fig:phantomResult} (d) & Average \\ \hline
	Manual 1 & 0.0509 & 0.0620 & 0.0565 & 0.0727 & 0.1196 & 0.0961 & 0.0665 & 0.0758 & 0.0711 & 0.0923 & 0.1450 & 0.1186 \\
	Manual 2 & 0.0595 & 0.0497 & 0.0546 & 0.1186 & 0.1034 & 0.1110 & 0.0709 & 0.0632 & 0.0671 & 0.1440 & 0.1237 & 0.1339 \\
	Mean of Manual & 0.0552 & 0.0559 & 0.0555 & 0.0957 & 0.1115 & 0.1036 & 0.0687 & 0.0695 & 0.0691 & 0.1182 & 0.1343 & 0.1263\\
	\textbf{PIPOC (Ours)} & \textbf{0.0193} & \textbf{0.0066} & \textbf{0.0130} & \textbf{0.0251} & \textbf{0.0200} & \textbf{0.0226} & \textbf{0.0220} & \textbf{0.0081} & \textbf{0.0150} & \textbf{0.0303} & \textbf{0.0245} & \textbf{0.0274}\\
	FIPOC \cite{ou2019automatic} & 0.0400 & - & 0.0400 & - & - & - & - & - & - & - & - & - \\ \toprule

\end{tabular*}
}
\label{tab:phantomExp}
\end{table*}

\begin{table}[!t]
\renewcommand{\arraystretch}{1}
\caption{False negative and false positive counts and ratios of joint location detection}
\centering
\begin{threeparttable}
\setlength{\tabcolsep}{1.09mm}{
\begin{tabular*}{\linewidth}{p{1.3cm}<{\centering}|p{1cm}<{\centering}p{1cm}<{\centering}p{1cm}<{\centering}p{1cm}<{\centering}|p{1cm}<{\centering}p{1cm}<{\centering}}
\bottomrule
	~ & \multicolumn{4}{c|}{False Negative} & \multicolumn{2}{c}{False positive} \\ \cline{2-7}
	 & IP & DIP & PIP & MCP & CMC & Others \\ \hline
	Thumb & 24 & N/A & N/A & 5 & 63 & 2 \\
	Index & N/A & 0 & 2 & 0 & 0 & 7 \\
	Middle & N/A & 1 & 1 & 0 & 0 & 2 \\
	Ring & N/A & 2 & 0 & 2 & 0 & 4 \\
	Small & N/A & 11 & 1 & 0 & 0 & 1 \\
	\textbf{Overall} & \tabincell{c}{\textbf{24}\\\textbf{(2.14\%)}} & \tabincell{c}{\textbf{14}\\\textbf{(0.31\%)}} & \tabincell{c}{\textbf{4}\\\textbf{(0.09\%)}} & \tabincell{c}{\textbf{7}\\\textbf{(0.16\%)}} & \textbf{63} & \textbf{16} \\ \toprule
\end{tabular*}
}
\begin{tablenotes}
\item[*] \textbf{IP}: Interphalangeal joint. \textbf{DIP}: Distal interphalangeal joint.\\ \textbf{PIP}: Proximal interphalangeal joint. \textbf{CMC}: Carpometacarpal joint. 
\end{tablenotes}
\label{tab:detectionResult}
\end{threeparttable}
\end{table}

\section{Experiments and Discussion}
\label{sec:experiment}

\subsection{Joint position detection}

The results of joint location detection are shown in Fig. \ref{fig:detectionResult}, the false negative of classifier decreases gradually with the increase of the cascade layers, and false positive gradually increase after $22$ levels.
We selected the $22$-layer classifier for joint position detection, which has a false negative ratio and a false positive ratio of $0.31\%$ and $0.50\%$ respectively. The performance on each joint is shown in the Table \ref{tab:detectionResult}. From this table, we can observe that false positives occurred mainly in the carpometacarpal (CMC) joint of the thumb, which is the joint that most closely resembles the target joints in a hand radiograph. And false negative appeared mainly in the thumb, especially the interphalangeal (IP) joint. In our opinion, the main reason for this situation is that the radiographic angle of the thumb is different from other fingers, resulting a difference in radiography. Differentiation of the joint position detection on the thumb may be effective in improving detection accuracy.

\subsection{JSN quantification}
\label{sec:JSNquantification}

\subsubsection{Phantom data}
\label{sec:phantomStudy}

Phantom images with ground truth were used in this experiment to calculate the absolute error of PIPOC, and to compare it with manual measurement. Manual measurement was done once with care by one radiologist and one radiological technologist after substantial training. They did not know the ground truth of the out-of-order phantom images. They were asked to determine the center of the upper phantom by drawing straight lines horizontally connecting both ends of the phantom base, then a straight line was drawn from the center vertically, and the JSW overlapping the straight line was measured.

Consider a set with $n$ phantom images. The mean error $E$ and root-mean-square deviation (RMSD) can be defined as:
\begin{equation}
E=\frac{2}{n(n-1)}\sum\nolimits^{n}_{f=2}(\sum\nolimits^{f-1}_{g=1}|\text{JSN}_{fg}-T_{fg}|)
\label{eq:error}
\end{equation}
\begin{equation}
\text{RMSD}\!=\!\sqrt{\frac{2}{n(n-1)}\sum\nolimits^{n}_{f=2}(\sum\nolimits^{f-1}_{g=1}(\text{JSN}_{fg}-T_{fg})^2)}
\label{eq:rmsd}
\end{equation}

Where $\text{JSN}_{fg}$ is the measured JSN between image $f$ and image $g$ by using PIPOC or manual measurement. And $T_{fg}$ represents the ground truth.

Figure \ref{fig:phantomResult} and Table \ref{tab:phantomExp} presents the measurement result of phantom sets. The manual measurement result of the radiologist and the radiological technologist showed high similarity in terms of mean error and RMSD in multiple phantom data sets.
The mean error of manual measurements is about $0.0555mm$ ($0.37pixel$) in low noise environment (air sets), and $0.1036mm$ ($0.69pixel$) in high noise environment (water sets).
This shows that visual measurement also can be greatly affected by the noise. On the other hand, this also indicates the manually annotated data have sub-pixel level mean error. Hence, the manually annotated ground truth may result in sub-pixel level deviation in algorithm evaluation of other works.

In paper \cite{ou2019automatic}, only one phantom set (environment: air, JSW range: $1.20mm$ - $2.20mm$, increment step size: $0.10mm$) is used in experiment.
The mean error of FIPOC is slightly lower than manual measurement. When compared to FIPOC, PIPOC can further improve the accuracy and robustness in JSN quantification, by eliminating the impact of image in-painting algorithm. As show in Table \ref{tab:phantomExp}, our work only has a $11.9\%$ to $35.0\%$ mean error, and a $11.7\%$ to $32.0\%$ RMSD when compare to manual measurement. This illustrates the improved performance of JSN quantification when using phantom datasets.

In related works, the ground truth of joint space is usually measured by the radiologist or the rheumatologist manually. But as discussed above, manual measurement also have sub-pixel level mean error. Thus, manually measured ground truth may result in sub-pixel deviation in algorithm evaluation. This deviation is negligible when evaluating the algorithm on a pixel scale. But it can be inaccurate on a sub-pixel scale.

To the best of our knowledge there are no published algorithms/methods which can compute ground truth RA joint space with sub-pixel accuracy. We propose to use the standard deviation $\sigma$ of multiple measurements to demonstrate the reliability of PIPOC without ground truth. The definition of standard deviation can be described as follows.

In case of three images $f$, $g$ and $k$, the $\text{JSN}_{fg-k}$ between image $f$ and image $g$ can be indirectly calculated by introducing intermediate image $k$. As show in Eq. \ref{eq:wijk}.
\begin{equation}
\text{JSN}_{fg-k}=\text{JSN}_{fk}+\text{JSN}_{kg}
\label{eq:wijk}
\end{equation}

Considering a set of images, the $\overline{\text{JSN}_{fg}}$ can be obtained by taking the average of multiple measurements.
\begin{equation}
\overline{\text{JSN}_{fg}}=\frac{1}{n}\sum\nolimits^{n}_{k=1}\text{JSN}_{fg-k}
\label{eq:wij}
\end{equation}

So, the standard deviation $\sigma$ of $\text{JSN}_{fg}$ is defined as Eq. \ref{eq:sd}.
\begin{equation}
\sigma_{fg}=\sqrt{\frac{1}{n}\sum^{n}_{k=1}(\text{JSN}_{fg-k}-\overline{\text{JSN}_{fg}})^2}
\label{eq:sd}
\end{equation}

\begin{figure}[!t]
\centering\includegraphics[width=\linewidth]{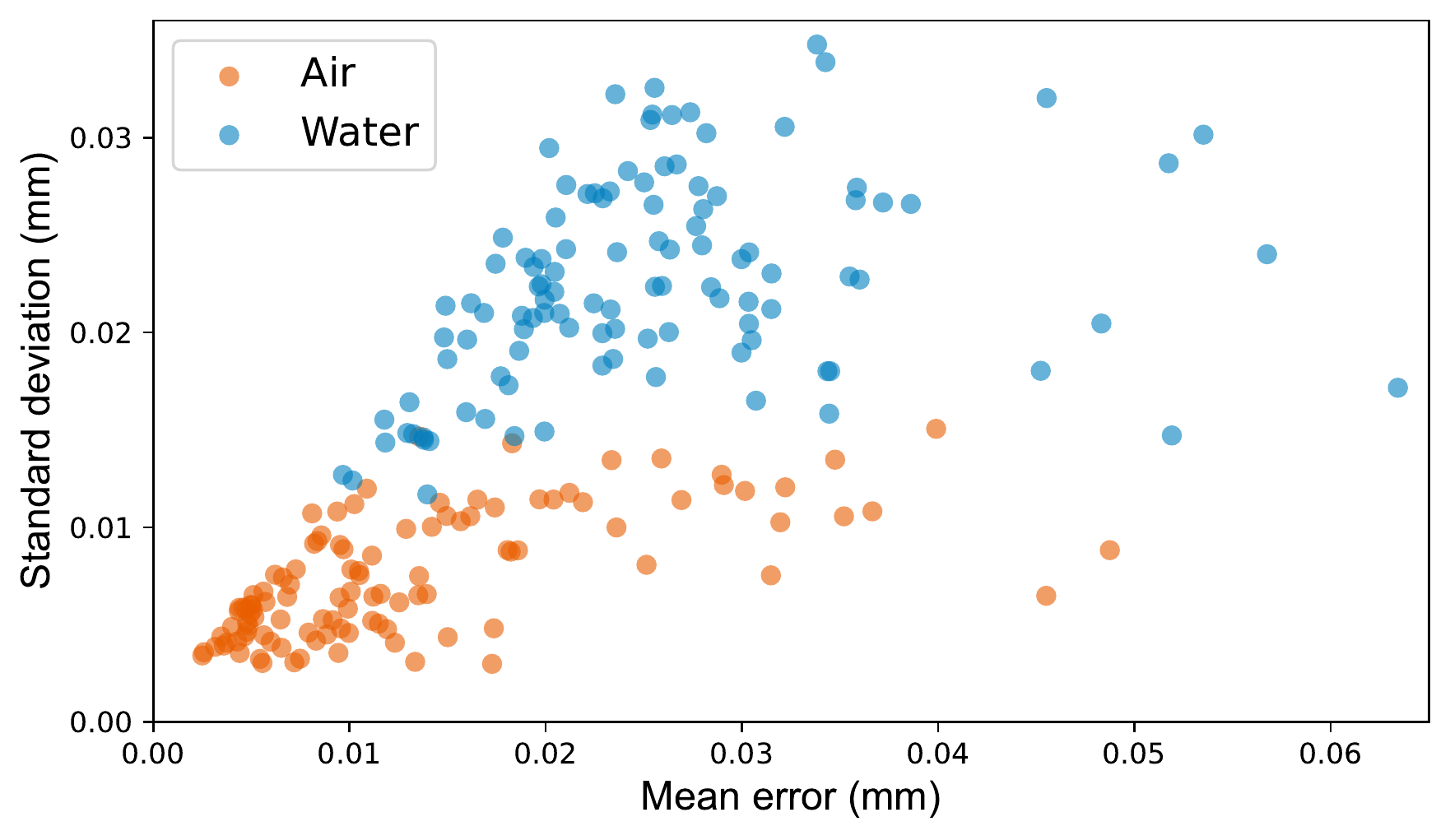}
\caption{The mean error $E$ and the standard deviation $\sigma$ of all sets of JSN in phantom data.}
\label{fig:SD_E}
\end{figure}

The standard deviation $\sigma_{fg}$ represents a dispersion of a set of $\text{JSN}_{fg-k}$ ($k\in[1,n]$). According to our experiments when using phantom datasets, the standard deviation $\sigma$ and the mean error $E$ has a high positive correlation, as show in Fig. \ref{fig:SD_E}. The Pearson correlation coefficient between $\sigma$ and $E$ is $0.641$ (count: $220$, $p$-value: $<.001$). For the above reason, and the most important advantage that the standard deviation $\sigma$ not relying on the ground truth, we used it to measure the performance of our work in clinical databases. In addition, we also found that noise in radiography due to beam attenuation in tissue can greatly affects the accuracy of measurements especially in terms of standard deviation.

\begin{figure}[!t]
\centering\includegraphics[width=\linewidth, trim=10 10 0 0, clip]{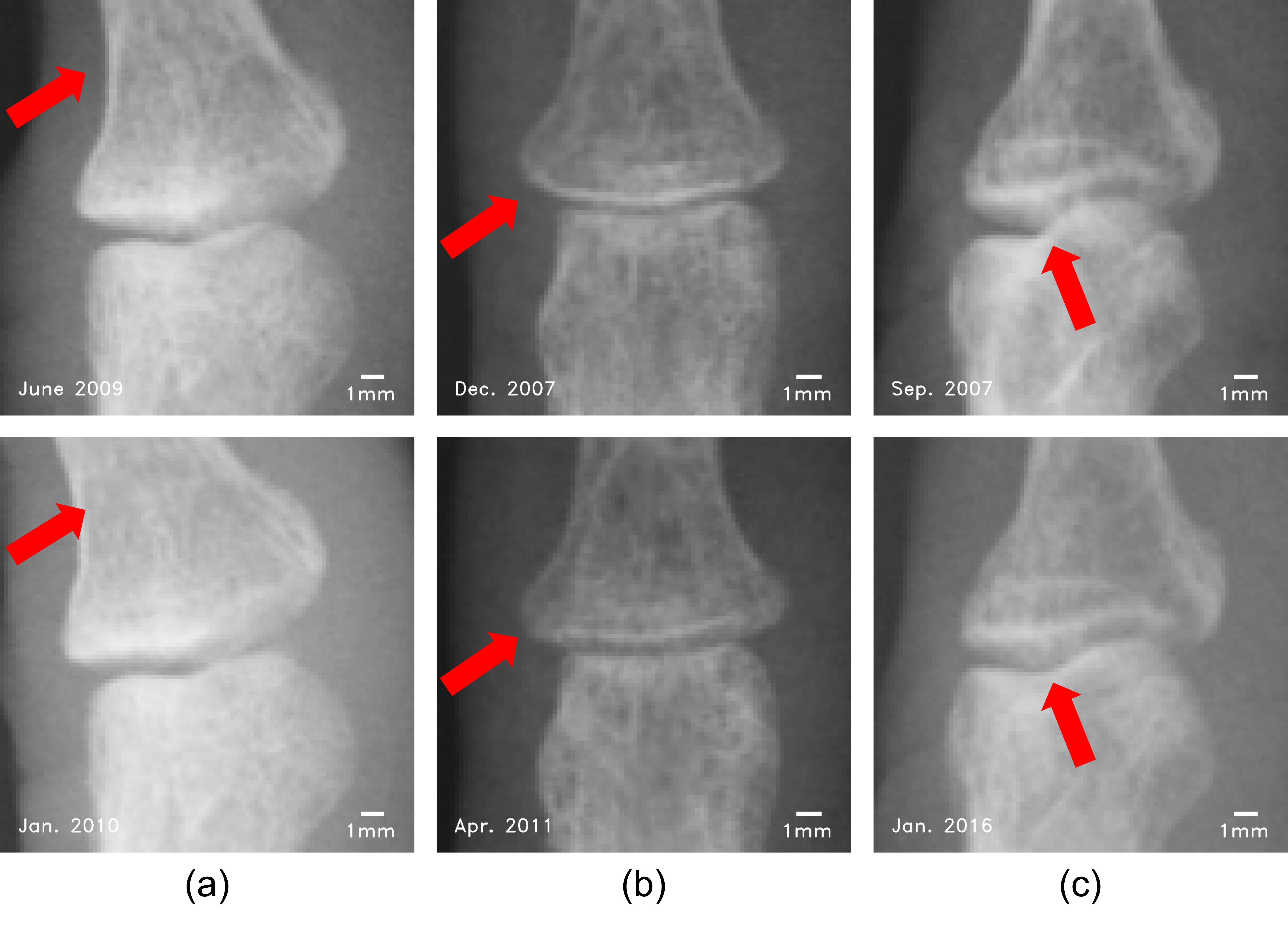}
\caption{Joints with mismatched registration. (a) Inconsistent joint angle. (b) Bended finger. (c) Inconsistent projection angle.}
\label{fig:mismatching}
\end{figure}

\begin{table}[!t]
\renewcommand{\arraystretch}{1}
\caption{The mean standard deviation in millimeter and the mismatching ratios for PIPOC}
\centering
\begin{threeparttable}
\setlength{\tabcolsep}{0.403mm}{
\begin{tabular*}{\linewidth}{c|cccc|cc}
\bottomrule
	~ & \multicolumn{4}{c|}{Clinical Data} & \multicolumn{2}{c}{Phantom Data} \\ \cline{2-7}
	~ & IP & DIP & PIP & MCP & Air & Water \\ \hline
	Thumb & 0.093(7.2\%) & N/A & N/A & 0.078(4.5\%) & - & - \\
	Index & N/A & 0.047(4.0\%) & 0.065(5.2\%) & 0.051(1.6\%) & - & - \\
	Middle & N/A & 0.055(5.8\%) & 0.061(3.4\%) & 0.057(4.3\%) & - & - \\
	Ring & N/A & 0.029(4.1\%) & 0.033(1.8\%) & 0.023(1.8\%) & - & - \\
	Small & N/A & 0.044(5.9\%) & 0.053(3.7\%) & 0.038(2.0\%) & - & - \\
	\textbf{Overall} & \textbf{0.093(7.2\%)} & \textbf{0.044(5.0\%)} & \textbf{0.053(3.5\%)} & \textbf{0.050(2.8\%)} & \textbf{0.007} & \textbf{0.025} \\ \toprule
\end{tabular*}
}
\label{tab:clinicalResult}
\end{threeparttable}
\end{table}

\begin{table*}[!t]
\renewcommand{\arraystretch}{1}
\caption{Comparison with related works. Mean error and standard deviation in millimeter. Numbers in braces indicate the corresponding percentage of the ground truth value for the respective joint.}
\setlength{\tabcolsep}{0.627mm}{
\begin{tabular*}{\linewidth}{lc|cc|cccc|cccc}
\bottomrule
	\multicolumn{2}{c|}{\multirow{2}{*}{Method}} & \multirow{2}{*}{Dataset} & Resolution & \multicolumn{4}{c|}{Mean Error} & \multicolumn{4}{c}{Standard Deviation} \\ \cline{5-12}
	\multicolumn{2}{c|}{} &  & (mm/pixel) & DIP & PIP & MCP & Overall & DIP & PIP & MCP & Overall \\ \hline
	Neural Network\cite{duryea1999automated} & AAPM'99 & 240 radiographs & 0.1 & 0.118 & 0.071 & 0.091 & 0.093 & - & - & - & - \\
	Active Shape Models\cite{langs2008automatic} & TMI'08 & 160 MCP joints & 0.0846 & - & - & 0.283(16.1\%) & 0.283(16.1\%) & - & - & 0.080(4.5\%) & 0.080(4.5\%) \\
	Edge Detection\cite{huo2015automatic} & TBME'15 & 104 radiographs & 0.1 & (5.8\%) & (7.2\%) & (7.1\%) & (6.8\%) & (4.8\%) & (5.3\%) & (4.4\%) & (4.8\%) \\
	Full Image POC\cite{ou2019automatic} & ISBI'19 & Phantom data & 0.15 & - & - & 0.040 & 0.040 & - & - & - & - \\
	Manual Measurement & - & Phantom data & 0.15 & - & - & 0.056 & 0.056 & - & - & - & - \\
	\textbf{Partial Image POC (Ours)} & - & Phantom data & 0.15 & - & - & \textbf{0.013} & \textbf{0.013} & - & - & \textbf{0.007} & \textbf{0.007} \\
	\textbf{Partial Image POC (Ours)} & - & 549 radiographs & 0.175 & - & - & - & - & \textbf{0.044} & \textbf{0.053} & \textbf{0.050} & \textbf{0.049} \\ \toprule
\end{tabular*}
}
\label{tab:related}
\end{table*}

\subsubsection{Clinical data}
$549$ hand PA radiographs have been analyzed in this subsection. Compared to phantom data, clinical data face additional challenges. The major challenge in this work is the uncertainty of hand posture, different hand postures can present differentiated bone contours. 

According to our experiments, changes in bone contours can affects the accuracy of JSN quantification. Here, we showcase (see Fig. \ref{fig:mismatching}) majority of mismatch bone contour cases. The most frequent reason is the inconsistent angle between the upper and lower bones of joint, as show in Fig. \ref{fig:mismatching} (a). This mainly occurs on IP and MCP joints. PIPOC has high accuracy for translation detection, but weak resistance to rotation.
Another important reason of mismatched registration is the bending of the fingers, which appears on DIP and PIP joints, for an example see Fig. \ref{fig:mismatching} (b). Finger bending can result in the changes of the far margin appearance of upper bone. Besides, inconsistent projection angle also can be the reason, see Fig. \ref{fig:mismatching} (c). Most of the time it happens only on the IP joint, which is caused by inconsistent joint position or thumb roll. 
The individuated finger movements differ greatly  as studied in \cite{hager2000quantifying}. Movements of the thumb, index finger, and little finger typically were more highly individuated than were movements of the middle or ring fingers. The angular motion tended to be greatest at the PIP joint of each digit \cite{hager2000quantifying}. It is worth noting that, the flexibility of joint and standard deviation express high positive correlation (refer Table \ref{tab:clinicalResult}).

In summary, the hand posture should be consistent and avoid bending of the fingers, especially the thumb when using our work for JSN quantification.
Thus, we strongly recommend that using guide lines lines to standardize hand posture in taking radiography, this simple step can greatly improve the accuracy of PIPOC.

\subsubsection{Comparison with related works}
Table \ref{tab:related} compares our work with previous JSW/JSN quantification works. In paper \cite{duryea1999automated}, they only used RMSD instead of mean error to evaluate the accuracy of their work, so we standardized the error metric accordingly. Considering that the error should conform to a Gaussian distribution, the mean error and RMSD can be transformed by Eq. \ref{eq:E_rmsd}.
\begin{equation}
\begin{aligned}
E&=\int^{+\infty}_{-\infty}|x|\cdot\frac{1}{\sqrt{2\pi}\cdot\text{RMSD}}e^{-\frac{x^2}{2\cdot\text{RMSD}^2}}dx\\
&=\sqrt{\frac{2}{\pi}}\cdot\text{RMSD}
\end{aligned}
\label{eq:E_rmsd}
\end{equation}

In paper \cite{huo2015automatic}, authors only give the corresponding percentage of the error to the ground truth. Considering the JSW of MCP is around $1.70mm$ \cite{kvien2006epidemiological, pfeil2007computer}, the mean error of MCP in millimeter is around $0.121$. It is noteworthy that, papers \cite{duryea1999automated, langs2008automatic, huo2015automatic} used manual measurement results as ground truth. As discussed above and in Table \ref{tab:related}, manual measurement has an error about $0.056mm$ (low noise) / $0.104mm$ (high noise) when using phantom data (spatial resolution: $0.15mm/pixel$). Although this value can decrease with higher spatial resolution, it is undeniable that in these works which employ manual measurement as the ground truth, the mean error may have a deviation.
 
The calculation procedure of standard deviation in paper \cite{langs2008automatic} is different from ours. 
They measured JSW of each joint $10$ times with varying clipping of the entire radiograph. The standard deviation quantified the uncertainty of measuring a radiograph.
In our work, an intermediate radiograph is introduced in standard deviation calculation.
The JSN progression between the two radiographs and the intermediate image is calculated respectively, thus, the standard deviation can be obtained by changing the intermediate image. 
When using the standard deviation calculation method given in paper \cite{langs2008automatic}, we measured a lower standard deviation (DIP: $0.0099mm$, PIP: $0.0095mm$, MCP: $0.0061mm$. These standard deviations do not include mismatched data, the mismatching ratios are shown in Table \ref{tab:clinicalResult}).

We can observe from Table \ref{tab:related} that even though the spatial resolution of our work is poorer than those in the related works, our mean error and the standard deviation are significantly lower.

\section{Conclusion and Future works}
\label{sec:conclusion}
Our work aims for computer-aided diagnosis of rheumatoid arthritis through automatic quantification of the joint space narrowing (JSN). We proposed an automatic algorithm pipeline, including joint position detection, joint segmentation and joint space narrowing quantification based on partial image phase-only correlation (PIPOC). From our experiments we observe highest JSN quantification accuracy when compared to similar algorithms. 

Additionally, our work can measure the displacements of upper and lower bones with sub-pixel accuracy. The measured mean error of our algorithm is in range of $11.9\%$ - $35.0\%$ in comparison to manual measurements using multiple phantom datasets, with a standard deviation of $0.0519mm$ when using clinical dataset. This algorithm greatly improves the accuracy and sensitivity of JSN quantification, which might help radiologists/rheumatologists to make more timely judgments on diagnosis and prognosis in rheumatoid arthritis patients.

Currently, machine learning is applied to difficult tasks, and we anticipate future studies in this direction. To address the posture (finger movement) related constraints and inconsistent joint angle which is likely to result in mismatched registration. We can quantify JSN by machine learning using the image features extracted by our work, this can improve the overall performance of the algorithm. 
 
\section{Acknowledgments}
The authors would like to sincerely thank Akira Sagawa, MD, Sagawa Akira Rheumatology Clinic (Sapporo, Japan),  Masaya Mukai, MD, Sapporo City General Hospital (Sapporo, Japan) and kazuhide Tanimura, MD, Hokkaido Medical Center for Rheumatic Diseases (Sapporo, Japan) for image data preparation.

\bibliographystyle{IEEEtran}
\bibliography{bibliography}

\end{document}